\documentclass[useAMS,usenatbib]{mn2e}
\usepackage{natbib}

\usepackage[dvips]{graphics}
\usepackage{epsfig}
\usepackage{amsmath}
\usepackage{aas_macros}

\def\mearth{{\rm\,M_\oplus}}

\def\deg{^\circ}
\def\micron{\mu {\rm m}}

\title[Mini Oort clouds from planet-planet scattering]{Mini-Oort clouds: Compact isotropic planetesimal clouds from planet-planet scattering}
\author[Raymond and Armitage]{Sean N. Raymond$^{1,2}$\thanks{E-mail: rayray.sean@gmail.com}
and Philip J. Armitage$^{3,4}$\\
$^{1}$CNRS, UMR 5804, Laboratoire d'Astrophysique de Bordeaux, 2 rue de l'Observatoire, BP 89, F-33271 Floirac Cedex, France\\
$^{2}$Universit{\'e} de Bordeaux, Observatoire Aquitain des Sciences de l'Univers, 2 rue de l'Observatoire, BP 89, F-33271 Floirac Cedex, France\\
$^{3}$JILA, University of Colorado \& NIST, 440 UCB, Boulder CO 80309-0440\\
$^{4}$Department of Astrophysical and Planetary Sciences, University of Colorado, Boulder}

\begin{document}

\date{Submitted Oct 11 2012 to MNRAS Letters}

\pagerange{\pageref{firstpage}--\pageref{lastpage}} \pubyear{2002}

\maketitle

\label{firstpage}

\begin{abstract}
Starting from planetary systems with three giant planets and an outer disk of planetesimals, we use dynamical simulations to show how dynamical instabilities can transform planetesimal disks into $10^2-10^3$ AU-scale isotropic clouds.  The instabilities involve a phase of planet-planet scattering that concludes with the ejection of one or more planets and the inward-scattering of the surviving gas giant(s) to remove them from direct dynamical contact with the planetesimals. ``Mini-Oort clouds" are thus formed from scattered planetesimals whose orbits are frozen by the abrupt disappearance of the perturbing giant planet.  Although the planetesimal orbits are virtually isotropic, the surviving giant planets tend to have modest inclinations (typically $\sim 10^\circ$) with respect to the initial orbital plane. The collisional lifetimes of mini-Oort clouds are long (10 Myr to $>$10 Gyr) and there is a window of $\sim$100 Myr or longer during which they produce spherical clouds of potentially observable dust at $70 \micron$. If the formation channel for hot Jupiters commonly involves planetary close encounters, we predict a correlation between this subset of extrasolar planetary systems and mini-Oort clouds.
\end{abstract}


\section{Introduction}

When gaseous protoplanetary disks disperse, many planetary systems may contain outer planetesimal disks with orbital
radii of $\sim 10-100$ AU. These discs are indirectly observable as debris disks~\citep{wyatt08,krivov10} if the
planetesimals are excited into an erosive collisional regime, whereby the largest bodies are slowly ground to dust
via collisional cascade~\citep{dohnanyi69}  The dust is detectable around other stars by observations of  far-IR
emission in excess of the stellar photosphere. About 16\% of Gyr-age Solar-type stars have debris disks that were
bright enough to be detectable at $70\micron$ with {\em Spitzer}~\citep{trilling08,carpenter09}; many more may
have fainter debris.  

The range of theoretically permissible planetesimal configurations is broader than the class of observed debris disks~\citep{heng10}. Which structures are realized in practice depends upon the formation and early dynamical evolution of planetesimal disks and planetary systems~\citep{kenyon08,krijt11,raymond11,raymond12}. In systems with giant planets, scattering events are likely to play a dominant role. 
At least 14\% and up to 50\% or more of stars host giant planets~\citep{cumming08,mayor11,gould10}.  The broad eccentricity distribution of giant exoplanets~\citep[median $\sim$0.25;][]{butler06,udry07b} is consistent with the idea that at least 50-75\% and up to $\sim$90\% of giant planet systems undergo dynamical instabilities involving planet-planet scattering events ~\citep{chatterjee08,juric08,raymond10}.  

Here we identify a novel dynamical route by which giant planet-planet scattering transforms the outer planetesimal disk into a cloud.  We call these ``mini-Oort clouds'' because the planetesimals' orbits are isotropic but on a spatial scale 100-1000 times smaller than the Solar System's $\sim 10^5$ AU-scale Oort cloud~\citep{oort50,weissman90}.  We discuss how the probability of forming isotropic planetesimal clouds depends upon the initial planetary system, and show that such structures may be common enough as to be present in samples of stars with unresolved IR excess emission.

\section{Simulations}
We identified mini-Oort clouds by mining a large database of simulations of planet-planet scattering in the presence of planetesimal disks, used for previous work on the planet-planet scattering mechanism and its consequences~(Raymond et al. 2008, 2009a, 2009b, 2010).  The initial conditions included a planetesimal disk extending from 10 to 20 AU containing $50 \mearth$ divided equally among 1000 planetesimals, which interacted gravitationally with the giant planets but not with each other.  The planetesimals followed an $r^{-1}$ surface density profile and were given initial eccentricities randomly chosen in the range 0-0.02 and inclinations between 0-1$\deg$.  The outermost giant planet was placed 2 Hill radii interior to the inner edge of the disk.  Two additional giant planets extended inward, separated by $\Delta$ mutual Hill radii $R_H,m$, where $\Delta = 4 - 5$ and $R_{H,m} = 1/2 (a_1+a_2) [(M_1+M_2)/3 M_\star]^{1/3}$ ($a$ refers to the planets' semimajor axes and $M$ to their masses, subscripts 1 and 2 to the individual planets, and $M_\star$ to the stellar mass).  In this suite of simulations we varied the giant planet masses and mass distribution~\citep[see][for details]{raymond10}.  The planets were given initially circular orbits with inclinations randomly chosen between zero and $1\deg$.  Each simulation was integrated for 100 Myr using the {\tt Mercury} hybrid symplectic integrator~\citep{chambers99} with a timestep of 20 days.  Collisions were treated as inelastic mergers conserving linear momentum.  Particles were considered to be ejected if they strayed more than 100 AU from the star.  Subsequently, we ran additional simulations with a much larger ejection radius to test the effect of this assumption (see Section 5).  

\section{Creation of a mini-Oort cloud}
\begin{figure}
  \begin{center} \leavevmode \epsfxsize=8cm\epsfbox{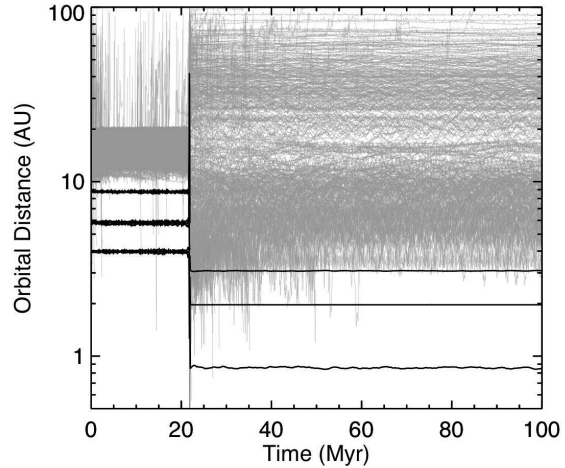}
    \caption[]{Evolution of a system that produced a mini-Oort cloud.  Plotted are the semimajor axis, perihelion and aphelion distances for the three giant planets in black, and the perihelion and aphelion distances for the planetesimals in gray.  The surviving giant planet has a semimarjor axis of 1.97 AU, an eccentricity of 0.57, and an inclination of $16^\circ$ with respect to the initial planetary plane.  An animation of the evolution of this simulation is available at http://www.obs.u-bordeaux1.fr/e3arths/raymond/movies/miniOC.mov .}  \label{fig:evol}
    \end{center}
\end{figure}

Figure~\ref{fig:evol} shows the formation of a mini Oort cloud in a system that started with three Jupiter-mass planets.  The planets remained on stable orbits for 21.7 Myr and then underwent a strong instability.  Planet-planet scattering caused the planets' orbits to rapidly spread out across the entire radial extent of the planetesimal disk.   After a series of a few hundred scattering events, the planet initially in the middle was ejected from the system at $t = 21.85$ Myr.  The two remaining planets underwent a further series of encounters that concluded with the ejection of the initially-outer planet and the inward-scattering of the initially-inner planet onto a long-term stable orbit with a semimajor axis of 1.97 AU, eccentricity of 0.57 and inclination of $16\deg$ with respect to the initial orbital plane.  

At early times the inner edge of the disk was eroded as planetesimals' eccentricities evolved to high enough values that their orbits crossed that of the outermost giant planet, usually leading to their ejection from the system.  When the giant planets became unstable and their orbits crossed the planetesimal disk, the giants began to scatter the planetesimals and systematically eject them from the system.  However, several close encounters are usually needed to impart enough orbital energy to eject a particle such that the clearing out phase takes some time.  In the simulation from Fig.~\ref{fig:evol}, the clearing out time exceeded the duration of the giant planet instability.  During the final planet-planet scattering phase, as one gas giant was systematically pushed outward on its way to being ejected, the other was pushed inward, closer to the star and away from the surviving planetesimals.  At the end of the instability the surviving planet was confined to the inner planetary system -- with an apocenter distance of just 3.1 AU -- and therefore dynamically separated from the planetesimals that survived the instability.  Those planetesimals were on unstable orbits in the process of being dynamically ejected, but were frozen in place when the giant planet perturbations stopped and their orbits suddenly became stable.  

\begin{figure}
  \begin{center} \leavevmode \epsfxsize=8cm\epsfbox{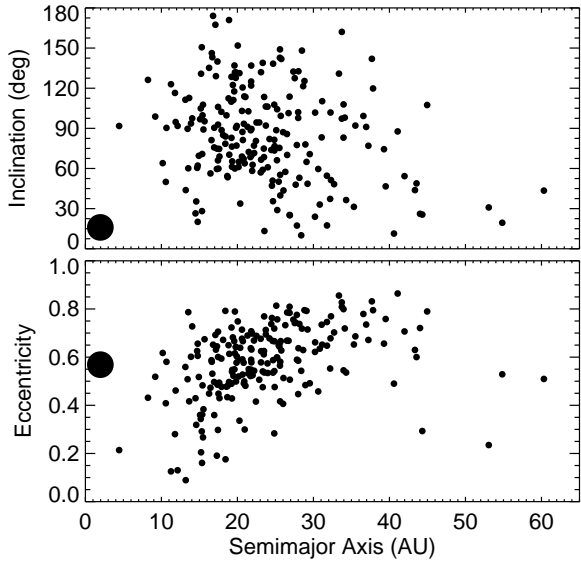}
    \caption[]{Final inclinations (top) and eccentricities (bottom) of the surviving particles in the simulation from Fig.~\ref{fig:evol}}  \label{fig:final}
    \end{center}
\end{figure}

How exactly did the planetesimals reach the isotropic inclination distribution seen in Figure~\ref{fig:final}?  During the instability the planets' inclinations were excited to up to $40\deg$.  The planetesimals that survived the instability  underwent multiple close encounters with at least one giant planet (with a median of 16 encounters).  The median closest encounter distance was $\sim 0.1$ AU, close enough to give the planetesimal a modest kick but in most cases not enough to eject it from the system.  Thus, the surviving planetesimals' orbits were stirred up by the giant planets but not strongly enough to eject them.  In other simulations not presented here the excitation level of surviving planetesimals covers the full range from planetesimals being only modestly perturbed by weak giant planet instabilities to all planetesimals being ejected by very strong instabilities.  Mini-Oort clouds represent an interesting middle ground where planetesimals are strongly excited but survive because the giant planets are dynamically removed on a short timescale.  

\section{Properties of mini-Oort cloud systems}

We found 36 mini-Oort cloud systems among the thousands of simulations we ran.  To be a mini-Oort cloud system we required that a) at least five planetesimal particles survived on stable orbits at the end of the simulation, b) that the median inclination of these planetesimals was $75\deg$ or higher, and c) that the system conserved energy well, specifically that the integrated d$E/E$ was less than $10^{-3}$.  

Mini-Oort clouds are preferentially produced in systems with $\sim$equal-mass giant planets. In the set of simulations contained three Jupiter-mass planets~\citep[referred to as ``jup\_disk'' in][]{raymond10}, 12 of 106 (11.3\%) unstable simulations with d$E/E<10^{-3}$ produced mini-Oort clouds.  This frequency was a factor of two lower in systems with equal-mass, Saturn-mass planets (12/265 = 4.5\% with mini-Oort clouds) and with equal-mass $3 M_J$ planets (3/57 = 5.3\%).  In the sets with randomly-chosen planet masses (following the mass distribution $dN/dM \sim M^{-1.1}$) the frequency of mini-Oort clouds was lower: 9/444 = 2\% for the set with masses chosen between $M_{Sat}$ and $3 M_J$~\citep[``mixed\_disk'' in][]{raymond10} and 2/531 = 0.4\% for the set with masses between $10 \mearth$ and $3 M_J$ (``mixed2\_disk'').  Nonetheless, the two planets closest in mass had a median mass ratio of just 1.13 among the 11 mini-Oort clouds in these mixed-mass simulations.  No mini-Oort clouds were found among the simulations in which planet masses were unequal, with mass ratios of $\sim 3$.  

The simulated mini-Oort clouds contained 5 to 213 planetesimals with a median of 27.  Given the simulated planetesimal mass of $0.05 \mearth$ these clouds contain $\sim 0.25-10 \mearth$ with a typical mass of $\sim 1 \mearth$ (although they evolve collisionally; see Section 6).  Mini-Oort cloud planetesimals have very excited orbits: the median (number-weighted) eccentricity was 0.53 and the median inclination was 85$\deg$.  The median planetesimal semimajor axis was $\sim 30$ AU.  The planetesimals' orbits undergo oscillations in eccentricity and inclination due to secular forcing from the giant planet.  In many cases these are in the Kozai regime and the oscillations have extremely large amplitudes.  This is clearly seen in the movie associated with Fig.~\ref{fig:evol}.  

\begin{figure}
  \begin{center} \leavevmode \epsfxsize=8cm\epsfbox{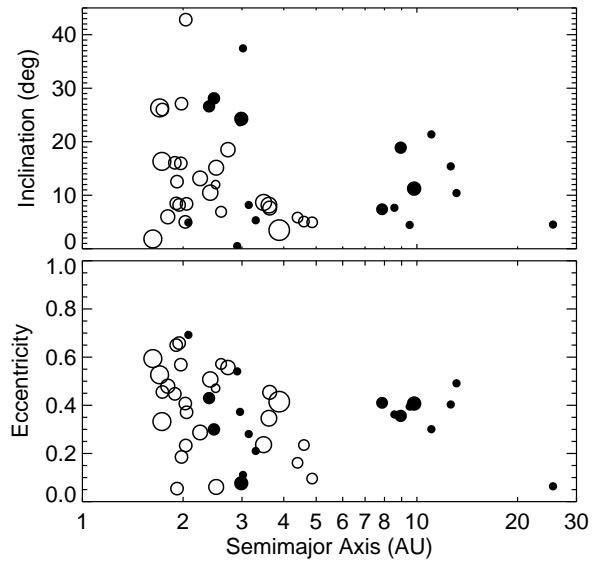}
    \caption[]{Orbital characteristics of the surviving giant planets in systems with mini-Oort clouds.  {\bf Top:} Inclination with respect to the starting plane of both the planets and planetesimal disk vs. semimajor axis.  {\bf Bottom:} Eccentricity vs. semimajor axis. In both panels systems with a single surviving planet are shown as black rings and systems with two surviving planets with solid circles.  The size of each circle is scaled to the giant planet's mass$^{1/3}$; for scale, the planets here range in mass from $1 M_{Sat}$ to $4.5 M_{Jup}$.}  \label{fig:aei_giants}
    \end{center}
\end{figure}

Figure~\ref{fig:aei_giants} shows the orbits of the surviving giant planet(s) in mini-Oort cloud systems.  The giant planets' large eccentricities serve as a reminder of the strong instabilities they underwent; the innermost planet has a median eccentricity of 0.41, substantially higher than the median of $\sim$0.25 for known giant exoplanets~\citep{butler06,udry07b}.  The inclinations of surviving planets with respect to the starting plane were modest, with typical values of $\sim10\deg$ and a maximum of just over $40\deg$, in stark contrast to the isotropic orbits of the planetesimal clouds.  

The bulk of mini-Oort clouds were produced in systems with a single surviving planet, but two giants survived in 9 of the 36 systems.  The clouds with two surviving giants were somewhat more distant and significantly lower-mass than with one surviving giant.  This comes about because with an additional, more distant giant planet and a fixed ejection radius there is much less dynamical room in which scattered planetesimals can survive.

\section{Effect of the simulation's ejection radius}
The major limitation of our large set of simulations is the 100 AU radius beyond which particles are considered to be ejected.  We ran 50 additional simulations with ejection radii of $10^{4-5}$ AU, typical values for the tidal stripping radius depending on the local stellar density~\citep[e.g.][]{tremaine93}.  These simulations had the same setup as before except for a smaller timestep (5 days) and a larger stellar radius (0.2 AU), both designed to minimize numerical errors.   All systems contained three $1 M_J$ planets in an attempt to maximize the production of mini-Oort clouds spaced by 4-5 mutual Hill radii in semimajor axis.  To accelerate the simulations we made the systems immediately unstable by giving the middle planet an eccentricity large enough to come within 1-2 Hill radii of another planet.  

In these runs, 8 of 36 (22\%) simulations with acceptable error tolerances produced mini-Oort clouds.  After 100 Myr, 6 additional simulations (17\%) contained mini-Oort clouds in the process of ejection, as the orbits of the isotropic planetesimals crossed those of surviving planets.  This confirms that mini-Oort clouds represent planetesimal disks in the process of ejection frozen in place when the giant planets are removed.  It also means that many planetary systems may undergo a phase during which a transient planetesimal population exists on isotropic orbits.  

The characteristics of the mini-Oort clouds in these simulations differ from the standard set in two interesting ways.  First, the mini-Oort clouds were more extended.  The number-weighted median planetesimal semimajor axis was 83.6 AU, and all of the mini-Oort clouds contained stable planetesimals at hundreds of AU and in some cases out to a few thousand AU.  The median (number-weighted) inclination of the surviving planetesimals was $88\deg$ and the median eccentricity was 0.7, consistent with an isotropic distribution of binary orbits.  Thus, the giant planets' repeated encounters with the planetesimals truly isotropized the planetesimal velocities, of course without exciting them to escape velocity.  

Second, some mini-Oort clouds were created in systems with no surviving giant planets.  This occurs when a very strong encounter between the last two surviving giant planets gives each high enough eccentricities to increase the apocenter distance of one beyond the ejection radius and decrease the pericenter distance of the other planet to below the stellar radius.  This occurred in three simulations, two of which produced mini-Oort clouds.  (All three of these simulations conserved energy to better than 2 parts in $10^5$.)  Three other mini-Oort clouds from this set were produced in systems where one giant planet collided with the star but an outer planet survived (at 8.2, 7.9, and 29.4 AU, similar to the clouds with 2 surviving giants in Fig.~\ref{fig:aei_giants}).  Most planets that are considered by the code to collide with the star should in reality survive on high-eccentricity orbits, as the simulated stellar radius was a factor of 40 larger than the Sun's.  If these planets' pericenter distances become small enough, tidal interactions with the star can act to re-circularize the orbits of very high-eccentricity planets and turn them into hot Jupiters~\citep{matsumura10,beauge12}.  Given that 5 of the 8 mini-Oort clouds in this set of simulations included a giant planet-star collision, a correlation may exist between hot Jupiters and mini-Oort clouds.  

These mini-Oort clouds contained roughly the same number of particles as in simulations with small ejection radii but were divided into two groups: 3 systems with $\sim200$ particles and 5 systems with 10-30 particles.  The low-mass clouds were preferentially found in systems with a relatively distant surviving giant planet, usually beyond 10 AU, but given the small number of high-mass clouds there was no clear correlation.  

\section{Collisional evolution and detectability of mini-Oort clouds}
To estimate the collisional evolution of mini-Oort clouds we assumed that each planetesimal represents an ensemble of smaller particles that are in collisional equilibrium and follow a $n(D) \sim D^{-3.5}$ distribution~\citep{dohnanyi69}, where $D$ is the particle diameter, assumed to be from $2.2\micron$ to 2000 km.  The collisional timescale $t_{coll}$ represents the typical time between catastrophic collisions for the largest (2000~km) bodies, calculated as in~\cite{raymond11,raymond12} following the equations laid out in previous work~\citep{wyatt99,wyatt07a,booth09}.  Although the planetesimal mass remained constant in our simulations, in our calculation of $t_{coll}$ we take into account the expected mass loss due to collisional grinding following the simple relation $M_{tot} = M_{init}/[1+t/t_{coll}]$~\citep{wyatt07a}.  From the orbital distribution of planetesimals and our knowledge of $t_{coll}$ we calculated the radial dust density and temperature distribution and the resulting dust flux, assuming the particles radiate as blackbodies~\citep[see Section 2 of][]{booth09}.

\begin{figure}
  \begin{center} \leavevmode \epsfxsize=8cm\epsfbox{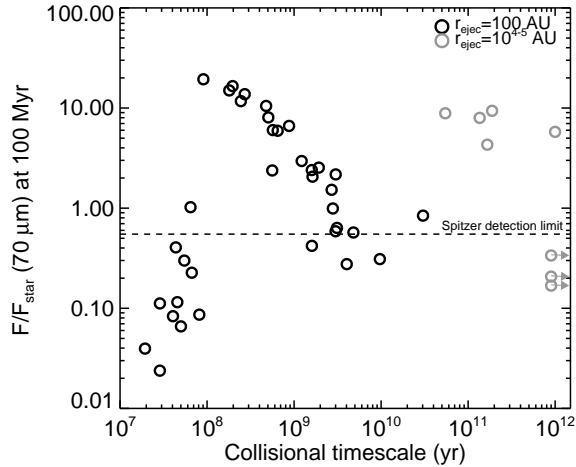}
    \caption[]{The dust-to-stellar flux ratio at $70 \micron$ as a function of collisional timescale $t_{coll}$ for the largest bodies (2000~km) bodies, calculated at the end of the simulations (100 Myr). The dashed line is the Spitzer detection limit~\citep{trilling08}.  Simulations with ejection radii $r_{ejec} = 100$~AU are in black and those with $r_{ejec} = 10^{4-5}$~AU in grey. } 
     \label{fig:tcoll}
    \end{center}
\end{figure}

Figure~\ref{fig:tcoll} shows the dust-to-stellar flux ratio at $\lambda=70 \micron$ as a function of $t_{coll}$ for each mini-Oort cloud after 100 Myr of dynamical and collisional evolution. The collisional timescales span from 10 Myr to far longer than the age of the Universe (as is the case for our own Oort cloud.  The simulations with ejection radii $r_{ejec} = 10^{4-5}$~AU have systematically higher $t_{coll}$ simply because they are more radially extended and $t_{coll}$ scales with the mean orbital radius of the planetesimals $R_m$ as $R_m^{2.5}$~\citep{wyatt07a}.  In systems with $r_{ejec} = 100$~AU with more than 50 surviving planetesimal particles $t_{coll} \approx 100-500$ Myr.  In most simulations $t_{coll}$ decreased sharply from a few hundred Myr to a few tens of Myr immediately after the instability, then increased quickly during the following $\sim10$~Myr.  

Mini-Oort clouds should be detectable by their dust emission, at least in their infancy.  After 100 Myr, 28 of 44 (64\%) of the clouds have dust fluxes above the Spitzer detection limit (Fig~\ref{fig:tcoll}).  In addition, 15 simulations (34\%) are above the Spitzer detection limit at $24 \micron$.  Even the simulations with low dust fluxes had detectable dust earlier in their evolution for an interval of at least 10 Myr.  The dust geometry should follow the orbital geometry and thus be an isotropic cloud.  To confirm the presence of a mini-Oort cloud thus requires information about the three dimensional structure of the dust, a significant observational challenge.  One potential method might be to infer dust along the line of sight by inspection of the stellar spectrum or colors at mid- to far-IR wavelengths~\citep[see][]{aumann84}.  

Mini-Oort clouds will grind down on the collisional timescale of the largest bodies, decreasing the dust and planetesimal masses and the corresponding dust flux (and thus increasing $t_{coll}$).  For all but a few simulations (those with $t_{coll} > 10^{11}$~yr) the collisional timescale for the smallest particles ($2.2\micron$) remains smaller than the timescale for P-R drag so our assumption of collisional equilibrium should hold, at least for the relatively short duration of the simulations~\citep[as is the case for known debris disks; see][]{dominik03,wyatt05b}.  On longer timescales, as the collisional times increase, P-R drag or corpuscular stellar wind drag may play a more prominent role in shaping the dust in mini-Oort clouds.  

\section{Summary}
We have shown that mini Oort clouds -- isotropic clouds of planetesimals with spatial scales of 100-1000 AU -- are a natural byproduct of planet-planet scattering in systems with nearly equal-mass, $\sim$Jupiter-mass planets.  These clouds are created by repeated close encounters with the scattered giant planets that isotropize the planetesimals' orbits with respect to the central star.  The outer giant planets are then removed by dynamical ejection or collision with the star (or perhaps by becoming hot Jupiters), leaving behind mini-Oort clouds on stable orbits.  The surviving giant planet(s) tend to have relatively compact ($< 10$ AU) high-eccentricity orbits with modest inclinations ($\sim 10\deg$; Fig~\ref{fig:aei_giants}).  In simulations with more realistic ejection radii of $10^{4-5}$ AU, 5 of 8 mini-Oort cloud systems included the collision of a giant planet with the star.  If star-planet tidal friction is efficient, this may suggest a correlation between mini-Oort clouds and hot Jupiters. 

Mini-Oort clouds may be detectable by their dust signatures at far-IR wavelengths.  More than half of our simulated clouds have detectable dust fluxes at $70\micron$ after 100 Myr, and there exists a window of 10-100 Myr or longer after the instability during which the dust from all mini-Oort clouds should be detectable. The collisional timescale for mini-Oort clouds is long (100 Myr to $>$10 Gyr) so the clouds should survive indefinitely albeit with decreasing dust production.  We note that there are several known systems with dust that may have high inclinations such as Lk Ca 15, HR 8799~\citep[see][]{krijt11}, although it is difficult to constrain the 3-D structure of dust clouds.

Finally, we note that there is an analogy between mini-Oort clouds and the Solar System's scattered disk~\citep{luu97,duncan97}.  Both contain small bodies on ``fossilized'' orbits and were produced during an era of planetary instability.  The modest inclinations in the scattered disk show that the Solar System's instability was far weaker than those typical of giant extra-solar planets.  

\section{Acknowledgments}
We thank Franck Hersant, Mark Wyatt, Philippe Thebault and Jean-Francois Lestrade for helpful discussions and the anonymous referee for a helpful report.  We thank Noel Gorelick and Google for the large amount of computer time donated for these simulations. S.N.R. acknowledges support from the CNRS's PNP program and NASA's Astrobiology Institute through the Virtual Planetary Laboratory lead team. P.J.A. acknowledges support from NASA under grant numbers NNX09AB90G and NNX11AE12G.


\end{document}